\documentclass[twocolumn]{aastex631}

\usepackage{graphicx}
\usepackage{amsmath}	
\usepackage{subfigure}
\usepackage{hyperref}
\usepackage{wrapfig}
\usepackage{booktabs}

\usepackage{comment}

\newcommand{\bz}{\langle B_z \rangle}
\newcommand{\nz}{\langle N_z \rangle}

\newcommand{\ha}{H$\alpha \;$}
\newcommand{\vsini}{$v\sin i$}
\newcommand{\vmac}{$v_{\rm mac}$}
\newcommand{\teff}{$T_{\rm eff}$}
\newcommand{\logg}{$\log g$}
\newcommand{\kms}{km~s$^{-1}$}
\newcommand{\stokesv}{Stokes~$V$}
\newcommand{\stokesi}{Stokes~$I$}
\newcommand{\esp}{ESPaDOnS}

\usepackage{xcolor}

\begin{document}


\shorttitle{The Complete Rotation Period of HD 54879}
\shortauthors{Erba, Folsom et al.}

\title{First Observation of the Complete Rotation Period of the Ultra-Slowly Rotating Magnetic O Star HD 54879}


\author{C. Erba}
\correspondingauthor{C. Erba}
\email{christi.erba@gmail.com}
\altaffiliation{co-first authors}
\affil{Department of Physics and Astronomy, East Tennessee State University, Johnson City, TN 37663, USA }

\author{C.~P. Folsom}
\altaffiliation{co-first authors}
\affil{Tartu Observatory, University of Tartu, Observatooriumi 1,T\~{o}ravere, 61602, Estonia}

\author{A. David-Uraz}
\affil{Department of Physics and Astronomy, Howard University, Washington, DC 20059, USA} 
\affil{Center for Research and Exploration in Space Science and Technology, and X-ray Astrophysics Laboratory, NASA/GSFC, Greenbelt, MD 20771, USA}

\author{G.~A. Wade}
\affil{Department of Physics and Space Science, Royal Military College of Canada, PO Box 17000, Kingston, ON K7K 7B4, Canada}

\author{S. Seadrow}
\affil{Department of Physics and Astronomy, and Bartol Research Institute, University of Delaware, Newark, DE 19716, USA}

\author{S. Bellotti}
\affil{Leiden Observatory, Leiden University, PO Box 9513, 2300 RA Leiden, The Netherlands}
\affil{Institut de Recherche en Astrophysique et Plan\'etologie, Universit\'e de Toulouse, CNRS, IRAP/UMR 5277, 14 avenue Edouard Belin, F-31400, Toulouse, France}

\author{L. Fossati}
\affil{Space Research Institute, Austrian Academy of Sciences, Schmiedlstrasse 6, A-8042 Graz, Austria}

\author{V. Petit}
\affil{Department of Physics and Astronomy, and Bartol Research Institute, University of Delaware, Newark, DE 19716, USA}

\author{M.~E. Shultz}
\affil{Department of Physics and Astronomy, and Bartol Research Institute, University of Delaware, Newark, DE 19716, USA}

\begin{abstract}
HD~54879 is the most recently discovered magnetic O-type star. Previous studies ruled out a rotation period shorter than 7 years, implying that HD~54879 is the second most slowly-rotating known magnetic O-type star. We report new high-resolution spectropolarimetric measurements of HD~54879, which confirm that a full stellar rotation cycle has been observed. We derive a stellar rotation period from the longitudinal magnetic field measurements of $\mathrm{P}=2562^{+63}_{-58}$~d (about 7.02~yr).
The radial velocity of HD~54879 has been stable over the last decade of observations. 
We explore equivalent widths and longitudinal magnetic fields calculated from lines of different elements, and conclude the atmosphere of HD~54879 is likely chemically homogeneous, with no strong evidence for chemical stratification or lateral abundance nonuniformities.
We present the first detailed magnetic map of the star, with an average surface magnetic field strength of 2954~G, and a strength for the dipole component of 3939 G. There is a significant amount of magnetic energy in the quadrupole components of the field (23 \%). 
Thus, we find HD~54879 has a strong magnetic field with a significantly complex topology.
\end{abstract}

\keywords{
Early-type stars (430),
Stellar magnetic fields (1610), 
Massive stars (732), 
Spectropolarimetry (1973), 
O stars (1137) 
}

\received{January 18, 2024}
\revised{August 19, 2024}
\accepted{October 11, 2024}


\section{Introduction} 
\label{sec:intro}
The large-scale spectropolarimetric surveys of OB stars conducted over the last two decades \citep[e.g., BoB, MiMeS;][]{Morel2015,Wade2016} have detected surface magnetic fields in around $\sim$7\% of massive OB stars. These fields typically have dipolar topologies, are globally organized, and are strong ($\sim$1~kG) and stable over timescales of decades \citep{Petit2013,Grunhut2017,Shultz2018a,Shultz2019d}.
Magnetic O-type stars are rare; thus, the detailed characterization of their field strengths and geometries is an important contribution to a developing understanding of this unique population.

The surface magnetic fields of OB stars significantly impact the star's long-term evolution (see, e.g., \citealt{Keszthelyi2023} for a recent review). \citet{Petit2017} and \citet{Georgy2017} used the mass-loss quenching formalism of \citet{udDoula2002} to demonstrate that magnetic massive stars retain much more mass during their main sequence evolution than inferred from non-magnetic evolutionary models. Similarly, the effect of magnetic braking, following \citet{udDoula2009}, has been incorporated into massive star evolutionary models, showing a rapid decline of the surface rotation \citep{Meynet2011,Keszthelyi2019,Keszthelyi2020,Keszthelyi2021,Keszthelyi2022}.
An extreme example of magnetic braking is found in the O4-7.5f?p star HD~108, which has the longest rotation period of all known O stars \citep[54~yrs;][]{Naze2001,Shultz2017a,Rauw2023}. HD~108 thus stands in stark contrast to the other 12 well-established magnetic O-type stars in our galaxy whose rotation periods are typically on the order of several days to months\footnote{The exception to this is HD~191612 (O6f?p–O8fp), with a rotation period of about 538~d \citep{Walborn2010,Wade2011}.} \citep{Petit2013,Grunhut2017}.
 
The topic of this study, HD~54879, is a main sequence O9.7V star \citep{Sota2011} that is distinguished by its strong magnetic field and very slow rotation. Spectropolarimetric measurements of the star were first obtained as a part of the B-fields in OB stars (BOB) survey \citep[e.g.,][]{Morel2015}, with the definite detection of the star's magnetic field initially announced by \citet{Castro2015} and \citet{Fossati2015}. Both papers reported the observed negative extremum of the star's longitudinal magnetic field ($\bz$) curve to be $\sim$~-600~G, implying a strong surface magnetic field of at least 2~kG. 
Additional spectropolarimetric monitoring by \citet{Wade2020b} and \citet{Jarvinen2022} showed a slow increase of the longitudinal field to an observed positive extremum of $\bz \sim$~74~G, before once again passing through magnetic null.

To date, estimates of the rotation period of HD~54879 have been based on fits to the incomplete monitoring of the longitudinal magnetic field curve. These indicated the rotation period was longer than five years \citep{Wade2020b}. The most recent sinusoidal fit reported by \citet{Jarvinen2022} suggested a stellar rotation period of 7.2~yr, placing HD~54879 as the second slowest rotator among the known magnetic O star population.

HD~54879's spectra show clear evidence of strong \ha emission, interpreted by \citet{Castro2015} as likely originating from circumstellar material. \citet{Shenar2017} reported variability in the star's \ha profiles over both short-term (days) and long-term (months to years) timescales, with the former being attributed to stochastic processes in the stellar wind \citep[e.g.,][]{Sundqvist2012,udDoula2013}, and the latter to variability associated with stellar rotation. These conclusions were supported with additional monitoring of the star reported by \citet{Hubrig2020} and \citet{Jarvinen2022}.  

In this paper, we report 
the most recent spectropolarimetric measurements of HD~54879. The data, described in Section~\ref{sec:obs}, indicate the $\bz$ curve has once again turned over, providing evidence that observations now span one complete rotation cycle. We perform a new magnetic analysis in Section~\ref{sec:mag}, and we provide a refined measurement of the stellar rotation period. We examine the radial velocity and \ha variability of the star, and investigate claims of differences in the strength of the longitudinal magnetic field of the star when the field is measured from lines of individual elements. We also present the first magnetic map of HD~54879. Our findings are discussed in Section~\ref{sec:concl}.

\section{Observations}
\label{sec:obs}

\subsection{\esp~Spectropolarimetry}
Three new sets of high-resolution ($R \sim 68,000$) spectropolarimetric sequences of HD~54879 were obtained using the \esp~echelle spectropolarimeter at the Canada-France-Hawaii Telescope (CFHT) on 2021 February 24, 2022 February 20, and 2023 October 21\footnote{Program codes 21AC15, 22AC28, and 23BC19.}. 
A log of these observations is provided in Table~\ref{tab:bz_obs}. The \esp~instrument covers an approximate wavelength range of 3700-10500 \AA~across 40 spectral orders. Each spectropolarimetic sequence consists of four subexposures, and yields four Stokes $I$ spectra, one Stokes $V$ spectrum, and two null ($N$) spectra, which were extracted using the methods described by \citet{Donati1997b}. A detailed description of the reduction
and analysis of \esp~data is provided in \citet{Wade2016}. 
Each subexposure in the 2021 and 2022 measurements had a duration of 867~s; similarly, each of the 2023 subexposures had a duration of 880~s. The peak signal-to-noise ratio per 1.8~\kms\ pixel near 550~nm of the recently obtained \stokesv~spectra ranges between 729 and 830. Thus, the new data are of similar quality to the available archival observations from \esp.

\subsection{Archival Data Sets}

In Table~\ref{tab:bz_obs} and in our corresponding analysis, we include all available archival spectropolarimetric observations of HD~54879 that were obtained with HARPSpol, mounted on the 3.6~m telescope at the European Southern Observatory’s (ESO) La Silla Observatory \citep{Piskunov2011}; with the Narval spectropolarimeter, mounted on the Bernard Lyot Telescope (TBL) at the Pic du Midi Observatory \citep{Auriere2003}; and with \esp~at CFHT. The original reduction and analysis of these data were presented by \citet{Castro2015}, \citet{Wade2020b} and \citet{Jarvinen2022}. We have performed a complete reanalysis of the archival data, in conjunction with the new observations, to provide as consistent and homogeneous a set of results as possible. Our extracted longitudinal magnetic field and null profile ($\nz$) values are consistent within uncertainties to the data reported by \citet{Wade2020b}, and are presented in Table~\ref{tab:bz_obs}. 


\section{Magnetic and Spectroscopic Analysis}
\label{sec:mag}

\begin{table*}
\centering
\caption{Archival and newly obtained (indicated by $^{\ast}$~in the Instrument column) spectropolarimetric observations of HD~54879, from HARPSpol (H), ESPaDOnS (E), and Narval (N). The longitudinal magnetic field ($\bz$) and null spectrum ($\nz$) are calculated using profiles computed with a least-squares deconvolution method, with an integration range of $+5$ to $+50$~km sec$^{-1}$. The phase $\phi$ is calculated using the ephemeris and 2562~d period described in Equation~\ref{eq:ephem}. The peak S/N is around 550~nm for all spectra.}  
\label{tab:bz_obs}
\begin{tabular}{lcccccl}
\hline \hline
HJD & $\langle B_z \rangle$ & $\langle N_z \rangle$ & RV & Peak & $\phi$ & Inst \\
- 2455000 & (G) & (G) & (\kms) & S/N  & &  \\
\hline
1770.4993 & -631$\pm$12 & -17$\pm$11 & 27.77$\pm$0.03 & 270 & 0.00 & H \\
1971.0658 & -561$\pm$12 & -1$\pm$10 & 27.64$\pm$0.06  & 795 & 0.08 & E \\
1971.1066 & -568$\pm$12 & 14$\pm$10 & 27.61$\pm$0.06 & 810 & 0.08 & E \\
2030.5195 & -542$\pm$19 & -4$\pm$18 & 27.74$\pm$0.06 & 459 & 0.10 & N \\
2032.5133 & -544$\pm$18 & -1$\pm$17 & 27.69$\pm$0.06 & 479 & 0.10 & N \\
2033.5159 & -531$\pm$19 & 1$\pm$17 & 27.76$\pm$0.06 & 480 & 0.10 & N \\
2092.5485 & -504$\pm$12 & -24$\pm$12 & 27.76$\pm$0.03 & 233 & 0.12 & H \\
2095.5057 & -509$\pm$19 & 15$\pm$19 & 27.76$\pm$0.03 & 156 & 0.13 & H \\
2310.6113 & -491$\pm$37 & 31$\pm$36 & 27.61$\pm$0.06 & 246 & 0.21 & N \\
2338.6680 & -445$\pm$38 & -10$\pm$37 & 27.60$\pm$0.06 & 229 & 0.22 & N \\
2357.6202 & -459$\pm$37 & 2$\pm$36 & 27.67$\pm$0.06 & 228 & 0.23 & N \\
2374.5860 & -444$\pm$33 & 16$\pm$33 & 27.69$\pm$0.06 & 255 & 0.23 & N \\
2409.5524 & -492$\pm$36 & -19$\pm$35 & 27.68$\pm$0.06 & 260 & 0.25 & N \\
2439.4167 & -538$\pm$49 & -72$\pm$49 & 27.56$\pm$0.06 & 191 & 0.26 & N \\
2736.9835 & -410$\pm$16 & -3$\pm$15 & 27.47$\pm$0.06 & 525 & 0.38 & E \\
2758.8779 & -409$\pm$17 & 6$\pm$16 & 27.46$\pm$0.06 & 518 & 0.38 & E \\
2775.9878 & -409$\pm$15 & -17$\pm$14 & 27.49$\pm$0.06 & 587 & 0.39 & E \\
2880.7390 & -348$\pm$17 & 12$\pm$17 & 27.61$\pm$0.06 & 553 & 0.43 & E \\
3008.1272 & -224$\pm$12 & -7$\pm$12 & 27.68$\pm$0.06 & 644 & 0.48 & E \\
3066.0351 & -182$\pm$12 & -5$\pm$12 & 27.55$\pm$0.06 & 649 & 0.50 & E \\
3128.9253 & -125$\pm$14 & 8$\pm$14 & 27.38$\pm$0.06 & 514 & 0.53 & E \\
3557.8753 & 75$\pm$15 & -2$\pm$15 & 27.58$\pm$0.06 & 658 & 0.70 & E \\
3744.1304 & -24$\pm$13 & 1$\pm$12 & 27.70$\pm$0.06 & 678  & 0.77 & E \\
4269.8770 & -637$\pm$14 & 2$\pm$12 & 27.54$\pm$0.06 & 729 & 0.97 & E$^{\ast}$ \\
4630.8714 & -500$\pm$11 & -3$\pm$9 & 27.53$\pm$0.06 & 830 & 1.12 & E$^{\ast}$ \\
5239.0541 & -427$\pm$11 & 0$\pm$10 & 27.57$\pm$0.05 & 779 & 1.35 & E$^{\ast}$ \\
\hline
\end{tabular}
\end{table*}

\begin{figure}
\centering
\includegraphics[width=\columnwidth]{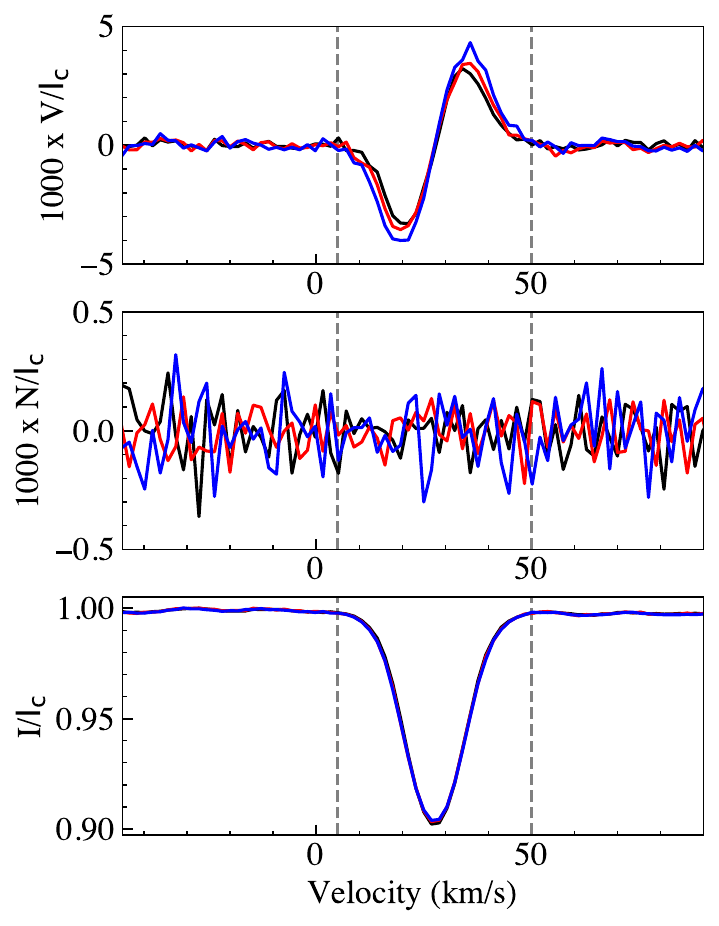}
\caption{Top to bottom: the \stokesv, null, and \stokesi~profiles extracted from the three most recent observations of HD~54879, corresponding to Julian dates ($- 2455000$) 4269.8770 (blue), 4630.8714 (red), and 5239.0541 (black) from Table~\ref{tab:bz_obs}. The integration range of $+5$ to $+50$~km sec$^{-1}$ is indicated by the gray dashed lines in each panel. The $V$ spectra deviate significantly from zero while the Null spectrum does not, indicating the detection of the star's magnetic field in these data. \label{fig:lsd_newdata} }
\end{figure}

\begin{figure}
\centering
\includegraphics[width=\columnwidth]{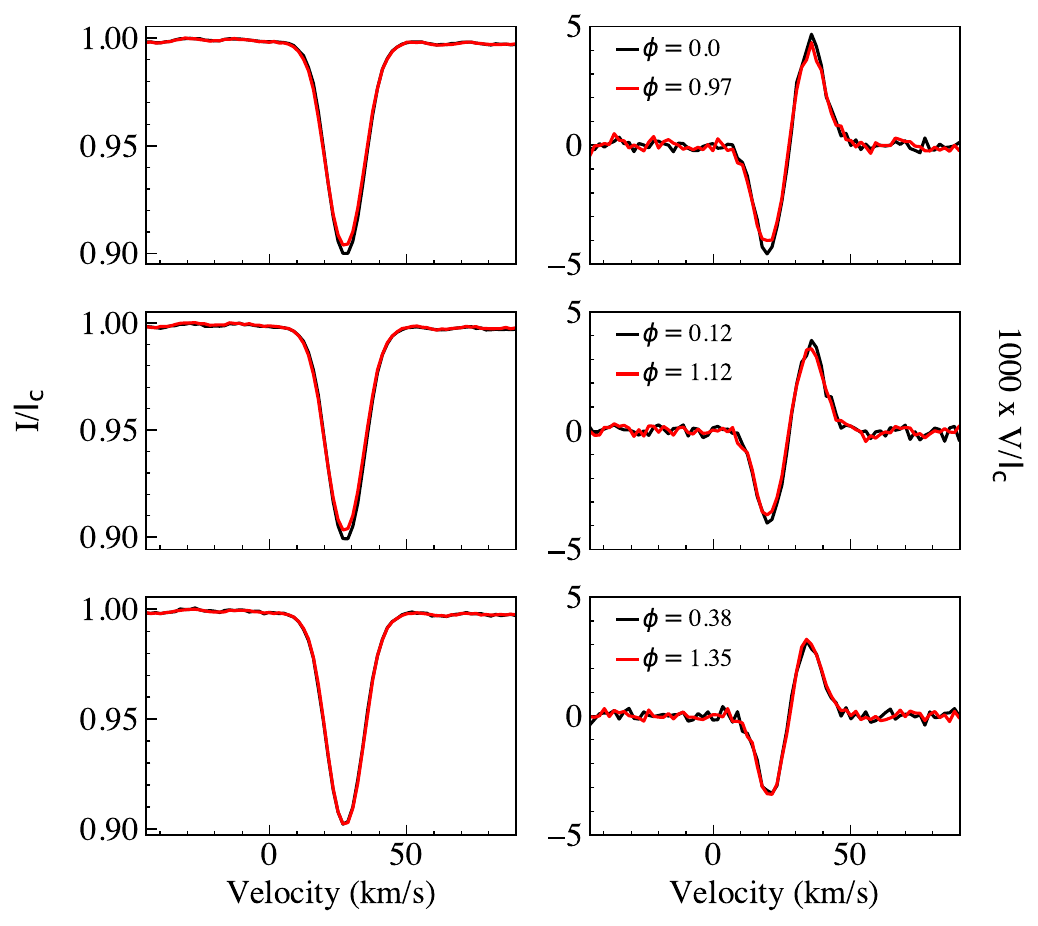}
\caption{\stokesi~(left column) and \stokesv~(right column) profiles of the most recent spectropolarimetric measurements of HD~54879 (red) compared to archival data of the star at approximately the same rotation phase $\phi$ (black), but one cycle previous to the new data. The profiles are consistent with each other. \label{fig:lsd_comp} }
\end{figure}

\begin{figure*}
\centering
\includegraphics[width=1.5\columnwidth]{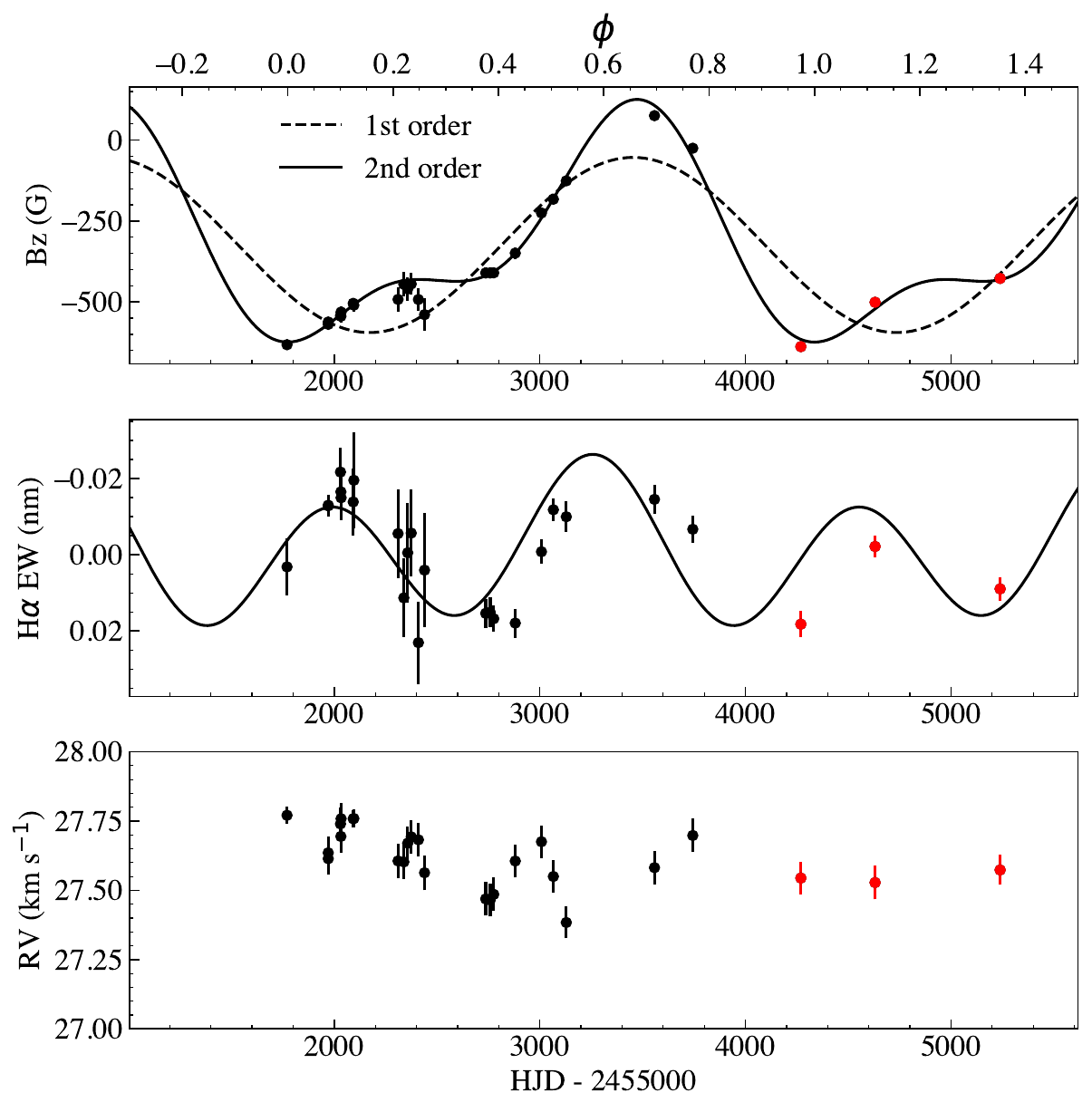}
\caption{{\bf Upper panel:} The phased time series of $\bz$ measurements. The lower x-axis references the date of the observations, with the archival data sets in black dots, and the data from the 2021-2023 observing epochs in red. The timestamps are phased according to an ephemeris with $P = 2562$~d and $t_0 = 2456774$ (HJD), reported along the top axis. The dashed and solid black lines show the 1st and 2nd order fits to the $\bz$ data, respectively.
{\bf Middle panel:} The phased time series of H$\alpha$ equivalent width measurements, with colors as above. Overplotted is the best fit using a 2-component harmonic function with the same period and ephemeris as the upper panel.
{\bf Lower panel:} Radial velocity measurements as a function of heliocentric Julian Date, with colors as above. \label{fig:bz_fit} }
\end{figure*}

\begin{figure*}
    \centering
    \includegraphics[width=0.9\textwidth]{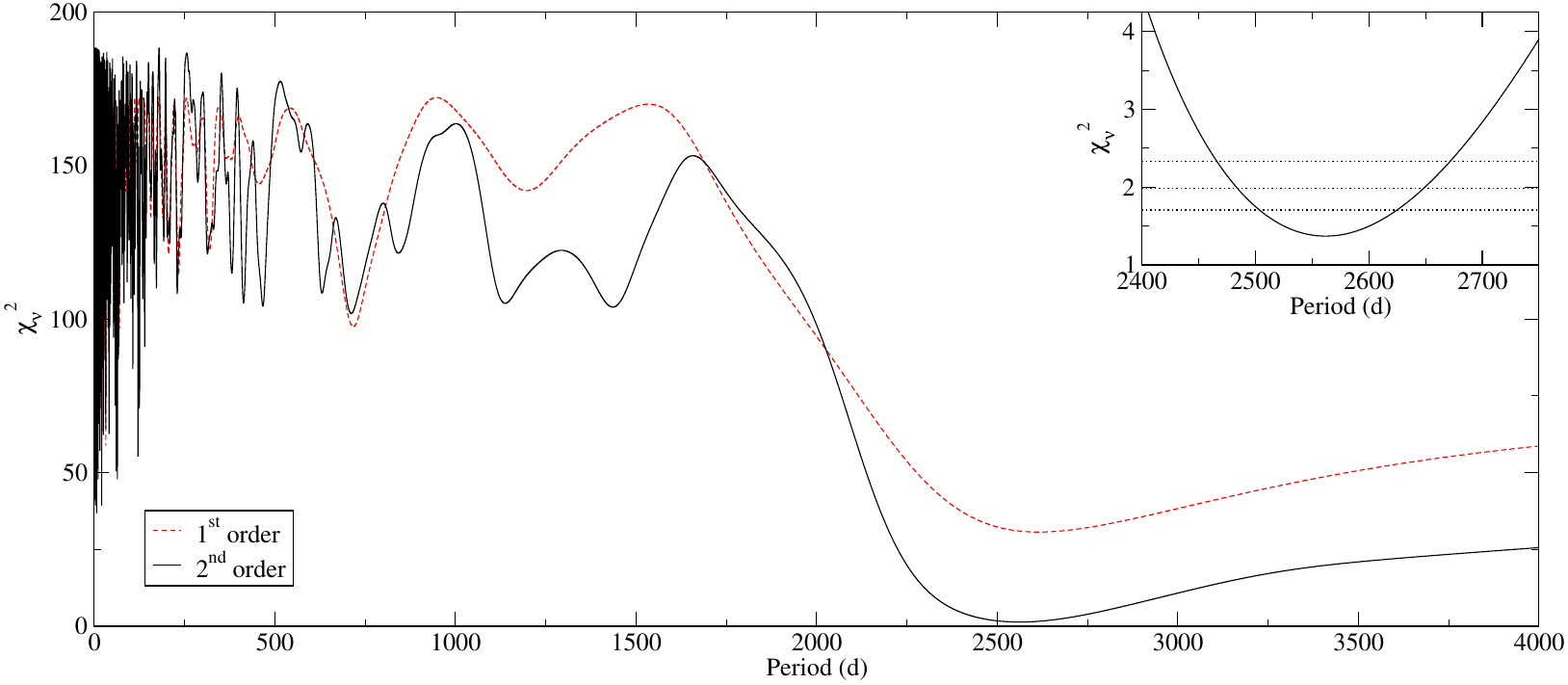}
\caption{Periodograms from fitting the $\bz$ data using 1st (red dashed line) or 2nd (black line) order Fourier series.  In the inset frame, dotted lines represent $\chi_\nu^2$ contours for 1, 2, and 3$\sigma$ confidence levels, about the best period for the 2nd order Fourier series fit. \label{fig:periodograms} }
\end{figure*}

\subsection{Longitudinal magnetic field}
\label{Longitudinal magnetic field}

The first step of the spectroscopic analysis was to normalize the spectra to their continuum. This was done by fitting low degree polynomials through carefully selected continuum points, and spectral orders were normalized individually.  
We started from unnormalized extracted spectra for both the new and archival observations, to provide a consistent normalization for all data.

We used Least-Squares Deconvolution (LSD; \citealt{Donati1997b}) to calculate pseudo-average line profiles for the intensity, \stokesv~and null spectra.  This technique effectively boosts the signal-to-noise ratio of the Zeeman signatures present in the \stokesv~profiles, to enable the detection of stellar magnetic fields. 
Here we use a new open source implementation of the LSD algorithm, \texttt{LSDpy}, contained within the \texttt{SpecpolFlow} software package \citep{spf-joss}. 
This follows the description of \citet{Kochukhov2010b}, and provides results fully consistent with their \texttt{iLSD} code.
We used the custom line mask of \citet{Wade2020b}, based on data from the Vienna Atomic Line Database (VALD3; \citealt{Piskunov1995, Ryabchikova1997, Ryabchikova2015, Kupka1999, Kupka2000}). This mask includes only metal lines, and excludes He lines, and uses lines with stable Stokes $I$ profiles to minimize any wind contribution. The observed He lines do not meet the self-similarity hypothesis of LSD, and display a variety of shapes due to varying amounts of Stark broadening, thus they were rejected.  We verified that this mask is well optimized for the spectral lines present in observations from all three instruments, and used it consistently for all the spectra. 
We used a normalizing wavelength of 500~nm, effective Land\'e factor of 1.2, and depth of 0.1 for the LSD calculation.
Every LSD \stokesv~profile produced a definite detection\footnote{The false alarm probabilities (FAP) were calculated using the following criteria: definite detection, FAP$<10^{-5}$; non-detection, FAP$>10^{-3}$ \citep{Donati1997b}.}, with the corresponding null profiles producing non-detections.  

The \stokesi, $V$, and null LSD profiles from the most recently obtained \esp~data are shown in Figure~\ref{fig:lsd_newdata}. 
The \stokesv~profiles show typical Zeeman signatures with a large amplitude (corresponding to the $\bz$ measurements in Table~\ref{tab:bz_obs}).  
In Figure~\ref{fig:lsd_comp}, we compare these profiles to archival observations obtained at similar rotation phases (using the ephemeris found below).  The profiles have nearly identical morphologies, amplitudes, and polarities, supporting the stability of the magnetic field and our derived period.

The longitudinal magnetic field was measured from the LSD profiles using a first moment technique \citep[e.g.,][]{Wade2000} with an integration range of $+5$ to $+50$~\kms, and the normalizing wavelength and Land\'e factor of the LSD profiles.  This is consistent with the procedure from \citet{Wade2020b}. We report the $\bz$ and $\nz$ (the $\bz$ calculation performed on the null profile) for all observations in Table \ref{tab:bz_obs}.
Measurements from the three new polarimetric sequences indicate a dramatic decrease in the magnitude of the longitudinal magnetic field (from $-24$~G to $-637$~G), with a subsequent, smaller increase (to $-500$~G and $-427$~G). These values are close to the initial measurements of the longitudinal magnetic field strength from \citet{Castro2015} and \citet{Wade2020b}, strongly suggesting the magnetic field has now been observed over one complete rotational cycle. 

Generally, the $\bz$ and $\nz$ values we find here are in good agreement with \citet{Wade2020b}.  Comparing results for the same observations, they mostly differ by $< 1\sigma$ of the joint error and always by $< 2\sigma$. 
Our results are in reasonable agreement with \citet{Castro2015}, who report $\bz$ for the first HARPSpol observation calculated with three different methods.  Compared with their `Bonn' LSD analysis (the most similar to our procedure) our result is stronger by -39 G with a joint error bar of 14 G, which is an acceptable agreement given differences in the line masks used.
Our results consistently disagree by $>3\sigma$ with the metal line analysis of \citet{Hubrig2020}, who reanalyzed the ESPaDOnS data of \citet{Wade2020b} but found differing $\bz$ values.  \citet{Hubrig2020} present results for two observations obtained on the same night (JD 2456971) with $\bz$ values differing by $\sim 5\sigma$ (81 G), which suggests they substantially underestimated the real uncertainty in their results. 
Our results disagree with \citet{Jarvinen2021} by $>3\sigma$ for about half of the observations we have in common, based on their LSD analysis using all metal lines (their most similar results).  \citet{Jarvinen2021} similarly disagree with \citet{Wade2020b}.  They partially agree with \citet{Hubrig2020}, with 6 of 9 of the measurements in common differing by $<2\sigma$.  \citet{Jarvinen2021} find large differences in $\bz$ over short periods of time (up to 196 G in 39 days), which are not seen in other studies of the same data (and not seen in some of their alternative line masks).  This suggests that there are significant systematic errors or underestimated uncertainties in some of their values.

The longitudinal magnetic field curve of HD~54879, presented in the upper panel of Figure~\ref{fig:bz_fit}, shows significant departures from what is expected from a dipolar topology (see also Section~\ref{sec:magMod} below). To date, topologically complex magnetic fields with strong departures from a dipole have only been reported in one other O-type star \citep[NGC~1624-2;][]{David-Uraz2021a,Jarvinen2021}, although complex geometries have also been found in a few early B-type stars \citep[e.g.,][]{Donati2006b, Kochukhov2011}. In this figure, 
we also show a first-order sinusoidal fit to the data (dashed line), which would be expected for a purely dipolar field.  Also included is a second-order harmonic fit to the data (solid line), which is equivalent to including both dipolar and quadrupolar components. 

Since the $\bz$ measurements now cover a full rotation cycle, we can perform a periodogram analaysis.
We proceeded by fitting Fourier functions to the data for a grid of fixed periods, and recording the resulting reduced chi-squared ($\chi_\nu^2$) for each period.  For a first order function this is equivalent to a Lomb-Scargle periodogram, except in units of $\chi_\nu^2$ rather than power, but this approach can be generalized for more complex magnetic fields by using a second or third order Fourier series.  Periodograms for both the first and second order Fourier series fits are shown in Figure~\ref{fig:periodograms}.  The first order fit produced a $\chi_\nu^2$ minimum near our adopted period, but with $\chi_\nu^2 > 30$, reflecting a strong departure from a dipolar magnetic field.  The second order fit reaches $\chi_\nu^2 = 1.37$, reflecting an adequate fit to the observations.  Using the second order fit, and placing the negative magnetic extremum at phase zero, we adopt the ephemeris:
\begin{equation}
    \label{eq:ephem}
    HJD = 2456774 + 2562^{+63}_{-58} E
\end{equation}
Thus the current best fit rotation period is $\mathrm{P}=2562^{+63}_{-58}$~d, or about 7.02~yr.

The departure of the $\bz$ curve from that expected for a pure dipole is highly statistically significant, which indicates a more complex magnetic topology. 
The success of the second order fit to the  $\bz$ curve suggests that there is a strong quadrupolar component to the magnetic field.
As noted above, such complex fields have been observed around cooler stars, such as $\tau$~Sco \citep[B0.2V;][]{Donati2006b}, $\alpha^2$~CVn \citep[B9p;][]{Silvester2014}, and HD~37776 \citep[B2V;][]{Kochukhov2011}, but had not yet been so prominently observed in the $\bz$ curve of an O-type star.  

\subsection{Radial velocity stability}
\label{Radial velocity stability}

With the LSD profiles calculated in Sect.~\ref{Longitudinal magnetic field}, we calculated precise radial velocities for HD 54879. The LSD calculations used a mask with most available metal lines in the spectra.  Radial velocities were determined by fitting a Gaussian to the Stokes $I$ profiles using $\chi^2$ minimization, and taking the centeroid.  Uncertainties were taken from the square root of the diagonal of the covariance matrix, and thus only incorporate the pixel uncertainties in the LSD Stokes $I$ profiles.

A plot of the radial velocity data as a function of time is in Figure \ref{fig:bz_fit} (lower panel), and this reveals no periodicity or significant trends with time.
The radial velocities are highly stable within observations from one instrument, and are in good agreement between instruments.  The full average is 27.62 \kms\ with a standard deviation of 0.10 \kms.  For just the 14 ESPaDOnS observations, the average is 27.56 \kms, with a mean uncertainty of 0.06 \kms, and standard deviation 0.08 \kms.  For the 9 Narval observations, the average is 27.67 \kms, with mean uncertainty and standard deviation of both 0.06 \kms.  For the HARPSpol data, the average is 27.76 \kms\ with a mean uncertainty of 0.03 \kms.  

We thus conclude that the radial velocity of HD 54879 has been stable to better than 0.1 \kms\ for the last 9.5 years. This definitively rules out proposals that this is an SB1 binary \citep[e.g.,][]{Hubrig2019}, unless the period is very long ($\gg 10$ yr).  This stability also argues against most other kinds of metal line variability, such as chemical spots, although we investigate that in more detail below.

\subsection{Measuring the magnetic field from lines of individual elements}
\label{sec:indelements}

\begin{figure}
\centering
\includegraphics[width=\columnwidth]{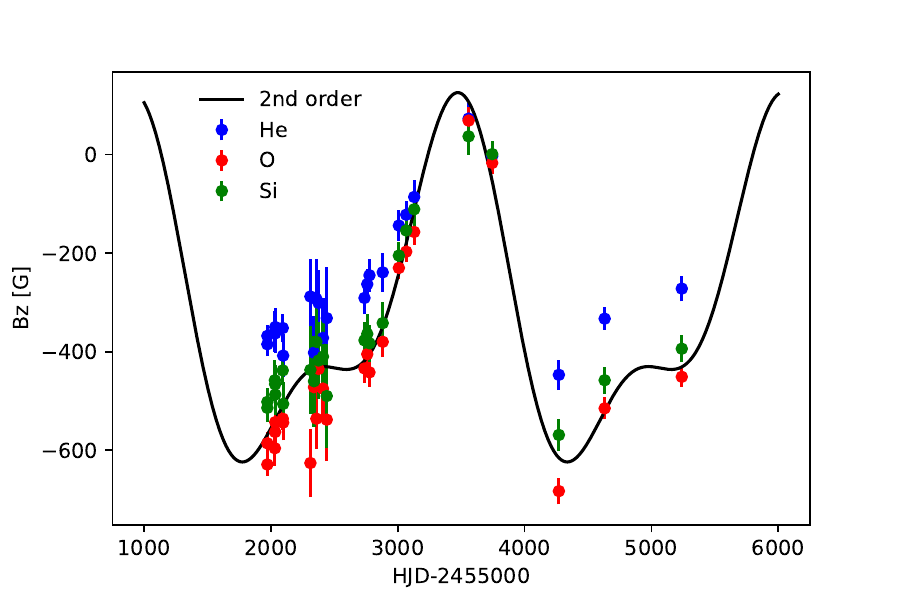}
\caption{Measurements of the longitudinal magnetic field from LSD profiles of  He, O and Si lines. The black  curve is the best-fit second-order sine curve from Section~\ref{sec:mag}.
\label{fig:bz_ind_element} }
\end{figure}

In their recent paper, \citet{Jarvinen2022} derive measurements of the longitudinal magnetic field strength of HD~54879 using single-element LSD line masks. They report that the $\bz$ measurements obtained from LSD profiles calculated using masks consisting individually of O, Si, and He lines are discrepant, while the LSD Stokes $I$ profiles do not show variability. The authors thus conclude that these results are reflective of ``different formation depths, with the He lines formed much higher in the stellar atmosphere compared to the silicon and the oxygen lines, and non-local thermodynamic equilibrium (NLTE) effects.'' This sentence may be implying that O, Si, and He have vertically non-uniform abundance distributions, i.e. that they are stratified. However, later in their article the authors appear to claim a nonuniform {\it surface} distribution of O, Si, and He: ``Thus, we conclude that the elements O, Si, and He are not uniformly distributed over the stellar surface of HD 54879...'' This latter sentence appears to imply that O, Si and He have non-uniform lateral distributions, possibly in addition to being vertically stratified.

The different LSD masks associated with each of the elements they analyzed are comprised of different collections of lines of different depths, land\'e factors, and widths, and certainly correspond to somewhat different measurement systems. Without calibration it is not surprising that the LSD profiles and longitudinal fields extracted using these different masks would be somewhat different. Nevertheless, we carried out an independent analysis of the observations using single-element line masks. We used the \texttt{iLSD} code of \citet{Kochukhov2010b} in multi-mask mode. (It is not clear which LSD code \citet{Jarvinen2022} employed.) Starting from the line mask of \citet{Wade2020b}, we created three new line masks containing only O lines, Si lines, and He lines, along with their associated anti-masks (i.e. the original mask excluding the lines of each of those elements). We then extracted LSD profiles for each element, taking into account the blending by other elements using the anti-masks. Finally, we measured the longitudinal magnetic field from each profile. Those measurements are shown in Figure~\ref{fig:bz_ind_element}.

\begin{figure}
\centering
\includegraphics[width=\columnwidth]{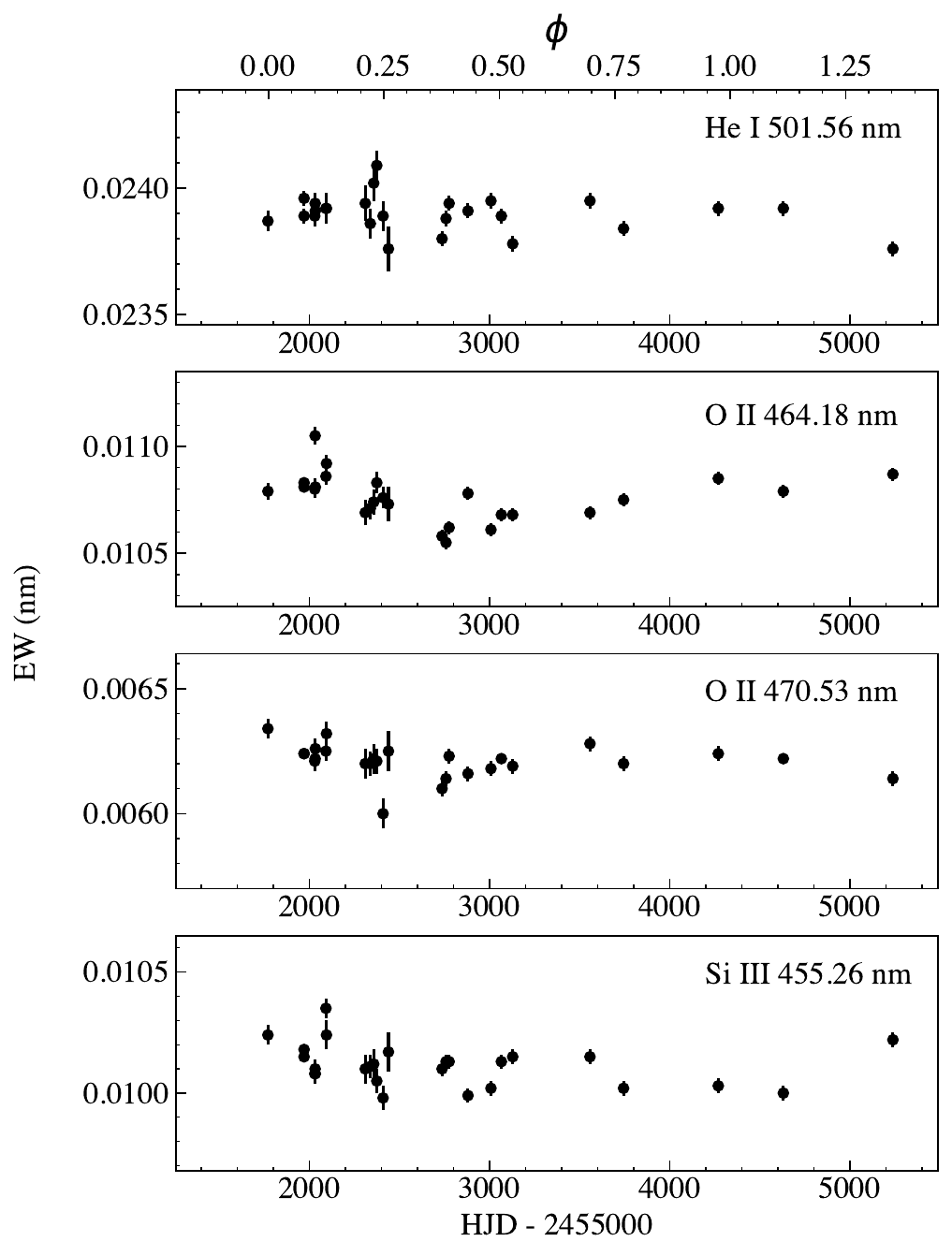}
\caption{Equivalent widths from individual lines of He, O and Si. No significant variability is evident.
\label{fig:singelem_ew} }
\end{figure}

As reported by \citet{Jarvinen2022}, there are some differences in the amplitudes of the $\bz$\ variations for individual elements. However, the various $\bz$\ curves have the same essential shapes, i.e. they are not strongly discrepant like those of chemically peculiar B-type stars such as HD\, 184927 \citep{2015MNRAS.447.1418Y}. Moreover, we observe no significant variations of line profiles or equivalent widths of these elements, either in LSD profiles or individual spectral lines. In Figure~\ref{fig:singelem_ew} we examine equivalent widths of lines chosen to be free from blending, strong enough to be clearly seen, but not so strong as to have large saturation effects, extreme NLTE effects, or a strong influence from the magnetosphere.
These lines of evidence all point to small differences in the measurement systems corresponding to the various line masks, leading to small differences in the inferred longitudinal fields\footnote{$\bz$ may be more sensitive to differences in the lines used for HD~54879 than for most stars.  The star has a strong magnetic field, with substantial cancellation in the line-of-sight component, which causes detectable Zeeman broadening (see Sect.~\ref{sec:magMod}). Thus differences between lines in Land\'e factor and splitting pattern may be more important here, especially for masks with few lines.}. 

In their papers, both \citet{Castro2015} and \citet{Wade2020b} show that the spectrum of HD~54879 is remarkably stable, especially for metal lines. This stability is contrasted by the variability of the \ha profile (see Section~\ref{sec:halpha}). These observations seem to indicate that HD~54879's dynamical magnetosphere does not demonstrably contribute in any significant way to the metal lines. Presumably, if there were a significant contribution from the dynamical magnetosphere, this would result in variability that would be visible in the equivalent width measurements, which as Figure~\ref{fig:singelem_ew} illustrates, is not observed. Moreover, the high resolution HARPS spectrum presented in \citet{Castro2015} is in good agreement with a FASTWIND synthetic spectrum corresponding to the inferred parameters of HD~54879, further supporting the assertion that most spectral lines are not significantly perturbed by the presence of the magnetosphere.
We thus conclude that there is no  compelling evidence for chemical stratification or lateral abundance nonuniformities in the atmosphere of HD~54879.

\subsection{H\texorpdfstring{$\alpha$}{a} Emission}
\label{sec:halpha}

\begin{figure*}
    \centering
    \includegraphics[trim={0.5cm 2cm 1cm 2cm},clip, width=\textwidth]{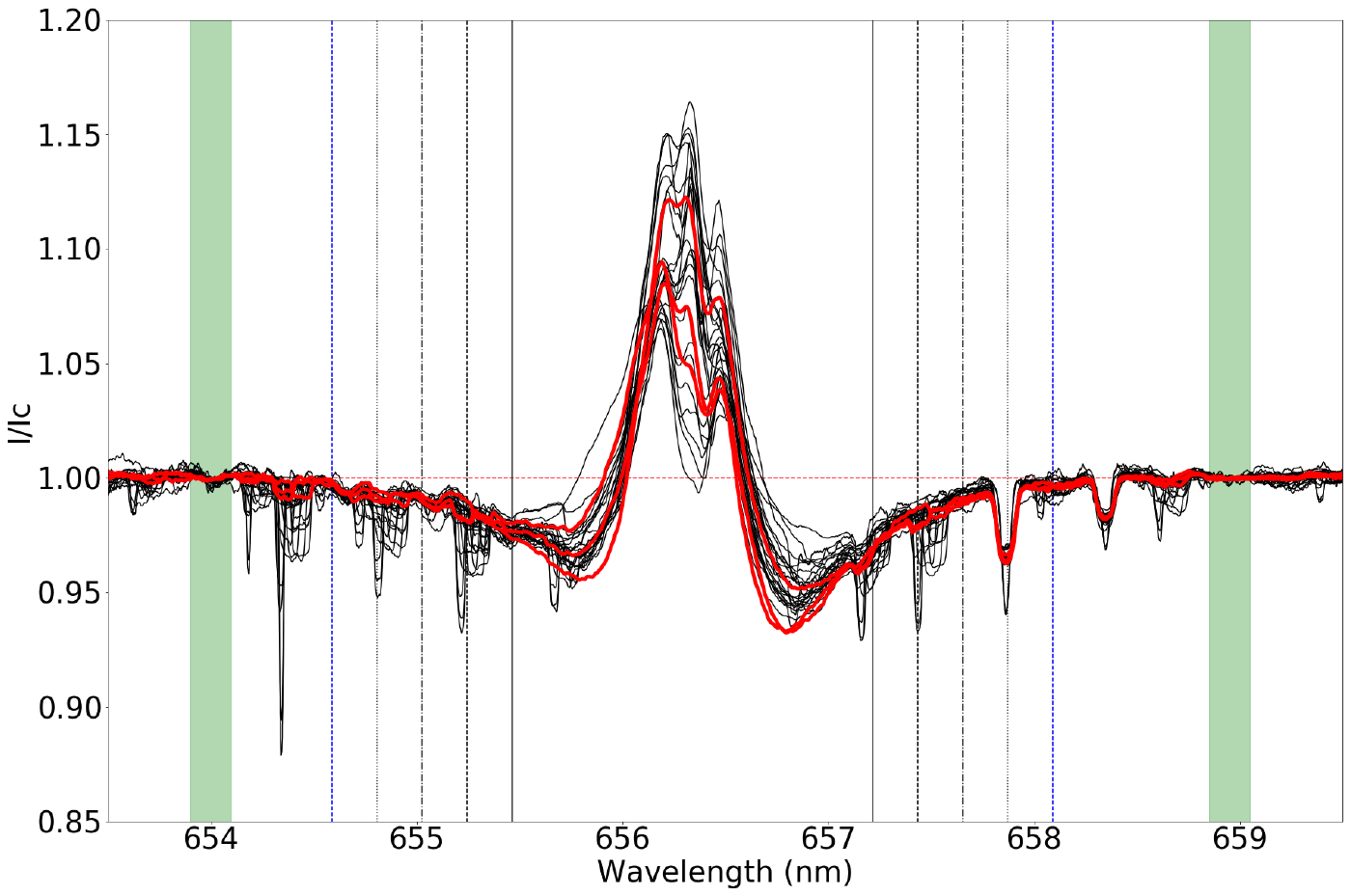}
    \caption{H$\alpha$ line from spectra taken with HARPS, NARVAL and ESPaDOnS (see Table~\ref{tab:bz_obs}). The archival data are plotted as black solid lines, while the new data sets are indicated in red. The continuum region used for local normalization are highlighted with green vertical shading. The other vertical lines correspond to the integration bounds that were tested for the calculation of equivalent widths, i.e. $\pm$400 km s$^{-1}$ (solid black), $\pm$500 km s$^{-1}$ (dashed black), $\pm$600 km s$^{-1}$ (dash-dotted black), $\pm$700 km s$^{-1}$ (dotted black), and $\pm$800 km s$^{-1}$ (dashed blue). We can see from this figure that HD~54879 exhibits significant H$\alpha$ emission at all rotational phases that were observed. \label{fig:halpha_overplot} }
\end{figure*}

HD~54879 exhibits atypically strong H$\alpha$ emission for its spectral type \citep{2007PASP..119..742B,Castro2015}, with variability on both short (days) and long (years) timescales \citep{Shenar2017}. Mass-loss rate estimates derived from this H$\alpha$ emission are comparatively too large \citep[by an order of magnitude;][]{Castro2015}, and so are considered unreliable\footnote{Mass-loss rate estimates for magnetic stars derived with spherically symmetric (non-magnetic) models will be inherently unreliable, since the magnetic channeling of the wind creates a fundamentally aspherical environment \citep[see e.g.,][]{Sundqvist2012,Marcolino2013,Erba2021b_uvadm}.}. \citet{Castro2015} suggested that HD~54879’s H$\alpha$ emission is likely magnetospheric in origin. For a star with a circumstellar magnetosphere, outflowing line-driven stellar winds can become trapped by the magnetic field and are forced to corotate with the star. Typically, in slowly rotating stars, rotation is not fast enough to provide sufficient centrifugal support for the plasma to overcome gravity. The confined wind material will fall back to the stellar surface under the influence of gravity on a dynamical timescale, forming a ``Dynamical Magnetosphere'' \citep[DM;][]{Sundqvist2012}. In the spectra of stars hosting DMs, H$\alpha$ emission has been observed to have a single broad peak close to line center \citep[e.g.,][]{Petit2013}. The multi-peaked H$\alpha$ emission from HD~54879, illustrated in Figure~\ref{fig:halpha_overplot}, suggests its circumstellar environment is shaped by a complex magnetic field. 

To explore the variation of the H$\alpha$ profile and how it relates to the rotation period derived from $\bz$ measurements, we compute equivalent width measurements using the spectra listed in Table~\ref{tab:bz_obs}. First, we refine the normalization that has already been applied to the spectra for the magnetometric analysis by defining continuum regions on either side of the H$\alpha$ line (653.9-654.1 nm blueward, and 658.85-659.05 nm redward), and renormalizing the H$\alpha$ line to a linear fit to those averaged regions. The resulting lines are shown overplotted in Figure \ref{fig:halpha_overplot}.

We then conducted some tests to determine the best integration range over which to compute our equivalent width measurements. We tried integration bounds at $\pm$400, 500, 600, 700, and 800 km s$^{-1}$ with respect to the H$\alpha$ line center (accounting for the systemic velocity of 27.62~km~s$^{-1}$; see Section~\ref{Radial velocity stability}).
As can be seen in Figure~\ref{fig:halpha_overplot}, all magnetospheric variability appears to be contained with the bounds at $\pm$400~km~s$^{-1}$, with larger bounds including undesirable photospheric lines overlapping with the wings of the H$\alpha$ line profile. Equivalent widths computed with these various bounds followed a similar trend from one spectrum to another, with uncertainties increasing with larger bounds. This indicates that indeed, the smallest bounds contain the bulk of the magnetospheric variability, and therefore they were used for the rest of the analysis (even though they do not span the full H$\alpha$ line profile).

Figure~\ref{fig:bz_fit} (middle panel)  
shows the clearly coherent variation of the H$\alpha$ equivalent widths with time. 
The measurements are fit with a second-order harmonic curve (black line) with a period corresponding to that derived from the $\bz$ analysis. 
Most of the equivalent width variability is consistent with this sinusoidal fit, modulated at the rotation period.  
However, we also performed an empirical harmonic fit to the \ha data with a partly unconstrained period (it was allowed to vary between 2400 and 2900~d). We found that a 2-component function is significantly preferred over a single sinusoid (with a reduced $\chi^2$ of 1.031 rather than 1.977).
This method (with the 2-component function) yielded a best-fit period of 2827~d ($\sim$7.75~yr), about 10\% longer than the stellar rotation period estimated from the best fit to the longitudinal magnetic field data. Minor discrepancies in estimates of the stellar rotation period derived from fitting \ha EWs and $\bz$ measurements are not unexpected, since the \ha emission can be modulated by processes other than the stellar rotation period \citep[see e.g., $\xi^1$~CMa, B0.5 IV;][]{Erba2021a_ksi1cma}. Future work will be needed to explore and interpret the modulation of the \ha emission in HD~54879's circumstellar environment. 

Shorter-term variability, as observed by \citet{Shenar2017} and also evidenced in our measurements (see Figure~\ref{fig:bz_fit}), could conceivably cause systematics in the determination of the period using equivalent width measurements, although it appears unlikely to fully explain the discrepancy between the periods derived from both methods. Further investigation, both observationally (more coverage) and theoretically (e.g., modeling H$\alpha$ emission using analytical prescriptions; \citealt{Owocki2016}) will be needed to resolve this tension. Nevertheless, the H$\alpha$ equivalent width variation supports our overall conclusion that HD~54879 rotates extremely slowly.

\subsection{Magnetic Geometry}
\label{sec:magMod}
The strong, complex field of HD~54879, combined with its long rotation period and assumed young age\footnote{In a standard picture, a very long rotation period implies the star has had time to experience spin down, and is therefore old. However, recent work investigating the effect of strong magnetic fields in main sequence stars \citep{Keszthelyi2019,Keszthelyi2020} and studying the outcomes of stellar merger events \citep{Schneider2019,Schneider2020} suggests there may be other pathways that produce young, slowly rotating magnetic massive stars. Complexity of the magnetic field could also be an indicator of stellar age, since higher-order multipolar components of the field decay first \citep{Charbonneau2013}.}, make it a particularly interesting candidate for Zeeman Doppler Imaging \citep[ZDI;][]{Donati1997a,Piskunov2002}, a technique used to create a surface map of the star’s  magnetic field. 
To date, only one O-type star has been mapped in detail using ZDI \citep[Plaskett's star;][]{Grunhut2022}.  The typically large macroturbulent broadening of O star spectral lines (and therefore Zeeman signatures) makes ZDI mapping challenging. 
HD~54879 has a relatively low macroturbulence (\vmac~$=8\pm3$ \kms, \citealt{Castro2015}; \vmac~$< 4$ \kms, \citealt{Shenar2017}; \vmac~$=8$ \kms, \citealt{Holgado2022}),
although its slow rotation implies \vsini\ is still much less than the macroturbulence.  Nevertheless, with the very high S/N and relatively complete rotation phase coverage provided by our observations, the most important large-scale components of the magnetic geometry can be reconstructed with ZDI.

For this analysis we adopted the rotation period ($2562^{+63}_{-58}$ d) and ephemeris from the $\bz$ data.  Using the rotation period and a radius one can derive the equatorial rotation velocity ($v_{\rm eq}$). \citet{Castro2015} reported $R = 6.7 ^{+1.0}_{-0.9}~R_\odot$ based on comparing spectroscopic \teff\ and \logg\ with evolutionary tracks from {\sc bonnsai}.
\citet{Shenar2017} reported $R = 6.1 \pm 1.5~R_\odot$, based on \teff\ and luminosity with the Stefan-Boltzmann law.  Here we adopted the value of \citet{Shenar2017}, although the exact value has very little impact on the ZDI results, giving us $v_{\rm eq} = 0.12 \pm 0.03$~\kms.

\subsubsection{Unno-Rachkovsky line model}

An initial attempt to fit the Stokes $V$ LSD profiles using ZDI and a line model relying on the weak field approximation \citep[specfically][]{Folsom2018a} failed.  The width of the Stokes $V$ profile varies and is narrower around phases 0.6-0.8, with the peaks closer to line center than most other phases.  With \vsini~$\leq 0.12$ \kms, a weak field model cannot reproduce this variation in Stokes $V$ width.  A model that incorporates Zeeman broadening due to a strong magnetic field is needed.

We used a modified version of the ZDIpy code
\citep{Folsom2018a}, that incorporates an Unno-Rachkovsky model for the local line profile. 
This code is described by \citet{Bellotti2023}, and follows the work of \citet{Donati2008_BPTau} and \citet{Morin2008}.
The code still uses a spherical harmonic description of the magnetic field \citep{Donati2006b}, the maximum entropy fitting routine of \citet{SkillingBryan1984}, and the entropy formulation of \citet{HobsonLasenby1998} calculated from the spherical harmonic coefficients.  
The Unno-Rachkovsky calculation follows the description in \citet{LandiLandolfi2004}. This is essentially a Milne-Eddington line model extended to solve the polarized radiative transfer equation.  Since we are modeling LSD profiles rather than individual lines, we assume a simple triplet Zeeman splitting pattern.  Local line opacity and anomalous dispersion profiles are taken to be Voigt and Faraday-Voigt profiles, respectively.  
\citet{Morin2008} and \citet{Bellotti2023} include filling factors in their line model, but these are not used here (or equivalently $f_I = f_V = 1$).

The model used here allows separate parameters for limb darkening (in continuum brightness), and the slope of the source function in the Milne-Eddington atmosphere ($\beta$ in equation [9.105] of \citealt{LandiLandolfi2004}).  This decouples the effects of brightness decreasing towards the limb and local line strength decreasing towards the limb.  In the calculations, the local line profiles are normalized by the local continuum and then scaled by the local limb darkening.  
However, the linear approximation of a Milne-Eddington atmosphere implies a linear limb darkening. The local continuum flux is
\begin{equation}
    I_c = B_0(1 + \beta \cos \theta)
\end{equation}
\citep[see][equations 9.105-9.112]{LandiLandolfi2004} $B_0$ is the surface value of the source function, and $\theta$ is the angle between the line of sight and surface normal. The local continuum flux relative to the flux at disk center is
\begin{equation}
    I_c/I_c^0 = (1 + \beta \cos \theta)/(1 + \beta).
\end{equation}
A standard linear limb darkening law \citep[e.g.,][]{Gray2005_Photospheres} is
\begin{equation}
    I_c/I_c^0 = 1 - \eta + \eta \cos \theta
\end{equation}
for a limb darkening coefficient $\eta$.  One can derive $\beta$ directly from $\eta$:
\begin{equation}
    \beta = \eta / (1 - \eta).
\end{equation}
We used this relation to set the slope of the source function to be consistent with the desired continuum limb darkening. This is convenient since tabulations of $\eta$ are available for a wide range of stars.

\subsubsection{Optimal model parameters}

\begin{figure*}
    \centering
    \includegraphics[width=0.9\linewidth]{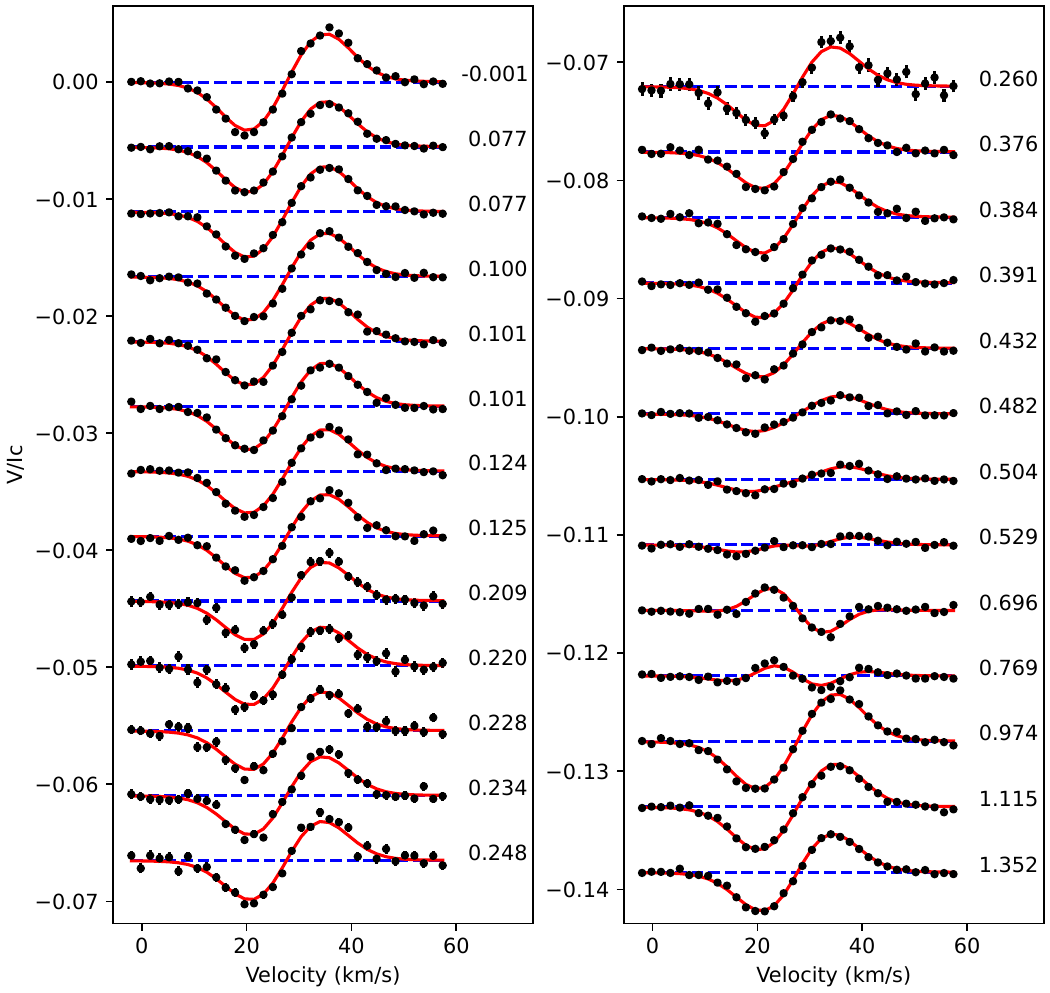}
    \caption{ZDI model fits (red) to the observed Stokes $V$ LSD profiles (black). Rotation cycles are labeled on the right.  Profiles are shifted vertically for clarity.  Error bars are plotted but are often smaller than the points. \label{fig:stokesV_fits} }
\end{figure*}

The ZDI model line requires a wavelength and effective Land\'e factor, which were set to the normalizing values for the LSD profile (500~nm and 1.2).  A limb darkening coefficient of 0.33 was used \citep{ClaretBloemen2011}, which implies a slope for the source function in the Unno-Rachkovsky model of 0.4925.  The local line opacity profile Gaussian width (combination of thermal and turbulent broadening), Lorentzian width (pressure broadening), and strength were fit using the Stokes $I$ and $V$ profiles.  Additional broadening by a Gaussian instrumental profile of $R = 65000$ was assumed.  The strong magnetic field produces a non-negligible amount of broadening in Stokes $I$, thus this fitting process was done iteratively with the magnetic mapping process described below.  The Gaussian (thermal + turbulent) broadening dominates the total line broadening, but is $\sim$1 \kms\ smaller in our final model than suggested by a non-magnetic model.  The final Gaussian width was 7.2~\kms\ ($\sqrt{2}\sigma$ of the Gaussian), as the midpoint between optimal values for Stokes $I$ and Stokes $V$.  This range leads to a potential systematic uncertainty of $\pm 0.5$~\kms, which spans the optimal values for Stokes $I$ and $V$. The Lorentzian width was 1.1~\kms\ ($\gamma$, half-width at half-maximum).  

The very low \vsini\ of HD~54879 limits the information that can be reconstructed with ZDI.  Smaller surface features, corresponding to higher degree spherical harmonics, are not resolved in the Stokes $V$ observations.  However, even lacking Doppler resolution, the few lowest degree spherical harmonics can be resolved by rotational modulation with dense phase coverage.  We restricted the maximum degree $l$ of the reconstructed spherical harmonics to $l_{\rm max} = 5$.  In testing a larger $l_{\rm max}$, the higher degree spherical harmonic coefficients tended to 0 due to the regularization.  Even modes with $l > 3$ had relatively small coefficients, suggesting we have little sensitivity to these modes. 
The toroidal components of the magnetic field are also virtually undetectable in Stokes $V$ at this \vsini.  At higher \vsini\ the toroidal components are mostly detectable in $V$ towards the limb of the star as regions of field with opposite orientation, but when \vsini~$=0$ these regions cancel out.  Consequently we restricted the magnetic geometry to a potential poloidal field.  In terms of the vector spherical harmonic coefficients\footnote{In this formalism $\alpha_{l,m}$ describes the radial field, $\beta_{l,m}$ describes the tangential polidal field, and $\gamma_{l,m}$ describes the tangential toroidal field.} defined by \citet{Donati2006b} (equations [2-4]), this restriction is equivalent to setting $\alpha_{l,m} = \beta_{l,m}$ and $\gamma_{l,m} = 0$.  If we allow the toroidal field to be free, it tends towards 0 due to the regularization; but this does not represent an absence of toroidal field, only our lack of sensitivity to this field component.  

The inclination of the rotation axis ($i$) is a significant uncertainty for this magnetic mapping.  We fixed $i$ by searching for the value that optimizes the ZDI result.  We ran a grid of ZDI models varying $i$ from $1^\circ$ to $90^\circ$ (in steps of $1^\circ$), with $v_{\rm eq} = 0.12$ \kms, and chose the model with the highest entropy that converged the target $\chi_\nu^2$ (set to 1.5).  To find a formal uncertainty on this inclination we repeated the grid search, but instead ran the code fitting to a target entropy rather than target $\chi_\nu^2$, with the target entropy set to the previous best value.  This produced a curve of $\chi_\nu^2$ as a function of $i$, for models with the same entropy, and the variation in $\chi_\nu^2$ about the minimum can be used to estimate an uncertainty.  This follows the procedure of \citet{PetitP2002}, discussed more by \citet{Folsom2018a} for $i$, and the approach was found to work well for cool stars with very low \vsini\ by \citet{Folsom2018b} and \citet{Folsom2020}.
This produced a best inclination of $i = 74 \pm 3 ^\circ$, although this may underestimate the real uncertainty due to the very high S/N of the LSD profiles.  Varying the Gaussian line width by $\pm0.5$ \kms\ varies $i$ by at most $1^\circ$, so this is a small contribution to the $i$ uncertainty.  Varying $v_{\rm eq}$ by its $\pm 0.03$~\kms\ uncertainty, varying the rotation period by its uncertainty, or changing the limb darkening coefficient by $\pm 0.1$, all change $i$ by less than $1^\circ$.

We used a similar grid search approach to verify the rotation period, following \citet{PetitP2002}.  Running a grid of ZDI models for a range of periods, we find $P = 2580 \pm 30$ d, which is in good agreement with the value from $\bz$ of $2562^{+63}_{-58}$ d.  Since the period from the ZDI search is more model dependent, and subject to more systematic uncertainties that are harder to quantify, we adopt the period from $\bz$ in the ZDI analysis and the rest of this study.

\subsubsection{The magnetic map}

\begin{figure}
    \centering
    \includegraphics[width=\linewidth]{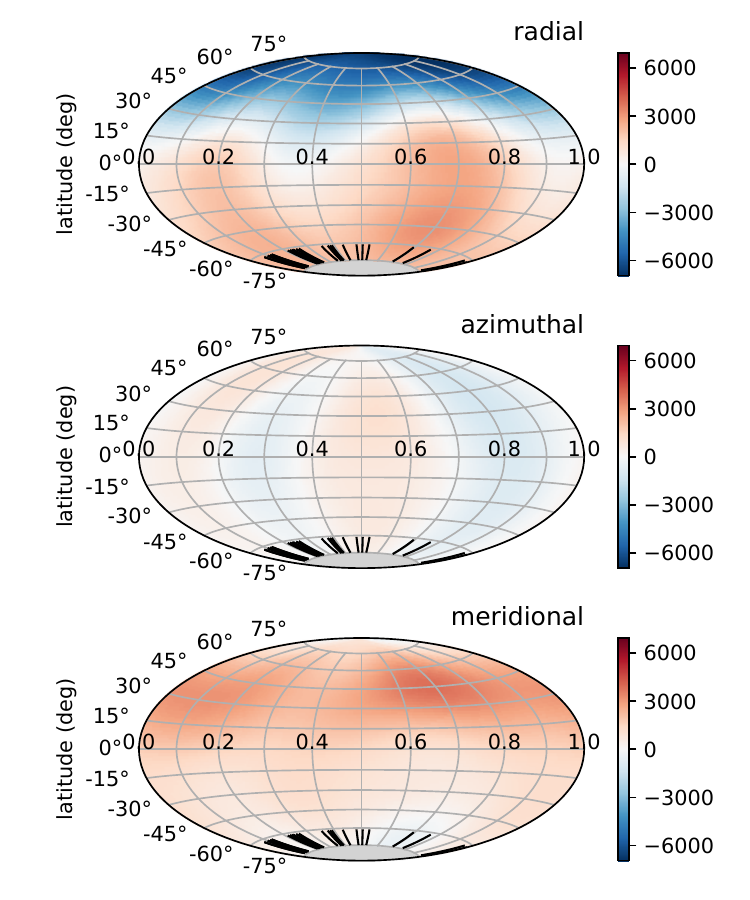}
    \caption{Magnetic map from ZDI for the radial, azimuthal, and meridional components of the magnetic field.  Presented in a Hammer projection with coordinates of latitude and rotation phase.  The color scale gives magnetic field in G.  Ticks at the bottom indicate phases of observations.  The grey area is the unobservable portion of the star. \label{fig:magnetic_map} }
\end{figure}

\begin{table}
    \centering
    \caption{Parameters describing the magnetic map.}
    \label{tab:magneic_geom}
    \begin{tabular}{lc}
    \hline\hline
    stellar parameter & value \\
    \hline
    P & $2562^{+63}_{-58}$ d \\
    \vsini & $0.116 \pm 0.029$ \kms \\
    $i$  & $74 \pm 3$\textdegree \\
    \hline
    magnetic parameter & value \\
    \hline
    mean $|B|$     & 2954 G\\
    maximum $|B|$  & 6961 G\\
    dipole $B_p$   & 3939 G\\
    dipole obliquity & 168\textdegree \\
    quadrupole $B_{\rm max}$ & 2299 G\\
    quadrupole $B_{\rm min}$ & -2331 G\\
    quadrupole max.\ locations  & 87\textdegree, 114\textdegree \\
    \,\,(colatitude, longitude) & 93\textdegree, 294\textdegree \\
    quadrupole min.\ locations  & 9\textdegree, 7\textdegree \\
    \,\,(colatitude, longitude) & 171\textdegree, 187\textdegree \\
    dipolar energy & 74\%\\
    quadrupolar energy & 23\%\\
    octupolar energy & 2\%\\
    total axisymmetry & 89\% \\
    dipolar axisymmetry & 96\% \\
    quadrupolar axisymmetry & 72\% \\
    \hline\hline
    \end{tabular} 
\end{table}

The final best fit to the Stokes $V$ LSD profiles is shown in Figure~\ref{fig:stokesV_fits}, and the corresponding magnetic map is shown in Figure~\ref{fig:magnetic_map}.  This result fits the data to $\chi_\nu^2 = 1.5$, which we consider acceptable given the small error bars in the LSD profiles.  
The surface average magnetic field strength is 2954 G,
reaching a maximum of 6961 G at a point 20$^\circ$ from the rotation axis near phase 0.  The dipole component has a polar strength of 3939 G and obliquity of the positive pole relative to the rotation axis of 168$^\circ$ at a longitude of 128\textdegree\ (phase 0.64; rotation phase runs in the opposite direction to longitude in this model). 
The magnetic geometry can be described in terms of the fraction of magnetic energy (proportional to $\oint B^2 \,d\Omega$) in different spherical harmonic components.  We find 74\% of the magnetic energy in the dipole ($l=1$) components, 23\% in the quadrupole (l=2) components and 2\% in the octupole (l=3) components.  The field is largely axisymmetric (about the rotation axis) with 89\% of the energy in axisymmetric ($m=0$) components.  
Parameters describing the magnetic field are presented in Table~\ref{tab:magneic_geom}.

The very low \vsini\ of HD~54879 leads to some large uncertainties in the magnetic map.  Most obviously, only the largest scale components of the magnetic field are reconstructed.  Components up to the octupole are resolved through the good rotation phase coverage, but beyond that (for $l > 3$) the geometry is largely unresolved.  The toroidal components of the magnetic field are similarly unresolved, due to the lack of Doppler resolution.  The toroidal field is typically weak in O and B stars \citep[e.g.,][]{Grunhut2022}, but in a few cases it is important \citep[][]{Donati2006b}.

The uncertainty on \vsini\ has a negligible impact on the magnetic field model, as do likely variations in the limb darkening coefficient.  Varying the rotation period by $\pm 1\sigma$ changes the magnetic field strength and geometry (energy fractions) by less than 1\%.  
Increasing the inclination by $1\sigma$ increases the magnetic field strength by $\sim$500 G and shifts 10\% of the energy into the dipole component.  Decreasing the inclination by the same amount decreases the magnetic field strength by $\sim$500 G but shifts 20\% of the dipolar energy to higher multipoles.
Increasing the Gaussian width by $0.5$ \kms\ increases the magnetic field strength by $\sim$100 G and shifts $\sim$3\% of the magnetic energy away from the dipole, and also causes the ZDI code to converge very slowly to the target $\chi_\nu^2$ of 1.5 (implying possible over-fitting, and that for this parameter combination a larger target $\chi_\nu^2$ likely should be chosen).  Decreasing the Gaussian width by $0.5$~\kms\ decreases the magnetic field strength by $\sim$30~G, and shifts the magnetic energy towards the dipole by $\sim$1\%. 

The variation in H$\alpha$ equivalent width correlates with the magnetic field map.  The maxima in H$\alpha$ emission occur near phases 0.1 and 0.6, which correspond to places where the positive radial field crosses above the equator, and maxima in the meridional field occur.  Minima in H$\alpha$ emission occur near phases 0.4 and 0.9, where the negative radial field spans the largest latitude range.  This correlation between surface magnetic field structure and H$\alpha$ emission strongly suggests that the stellar wind is being influenced by the magnetic field.  Material from the wind can become trapped in closed magnetic field loops above the surface of the star, leading to enhanced H$\alpha$ emission.  HD~54879 has been noted to have an unusually strong H$\alpha$ emission for its spectral type \citep{2007PASP..119..742B, Castro2015}.  This could be explained by the very strong axisymmetric dipolar component of the magnetic field, which should lead to large closed field loops being observable at all rotation phases. 
Modeling of the magnetosphere around the star is needed to fully explore this hypothesis, such as the Analytic Dynamical Magnetosphere approach \citep{Owocki2016, Munoz2020} or MHD simulations.

\section{Discussion and Conclusions}
\label{sec:concl}

The most recent spectropolarimetric data of HD~54879 indicate that the longitudinal magnetic field has once again reached the negative magnetic extremum and begun increasing, implying the full stellar rotation period has now been observed. Our best empirical fit to the longitudinal field data yields a stellar rotation period of $\mathrm{P}=2562^{+63}_{-58}$~d, which is about 7.02~yr. 

We revisited the longitudinal magnetic field variation measured from LSD profiles constructed from lines of individual elements, specifically He, O and Si. While those longitudinal field curves show some small differences in amplitude, their similar shapes, compounded by the lack of any significant line profile or EW variability of these elements, leads us to attribute these discrepancies to small differences in the measurement systems associated with the different line masks. In contrast to \citet{Jarvinen2022}, we see no evidence for chemical stratification or lateral abundance nonuniformities.

Unlike the majority of other known magnetic O stars, HD~54879's longitudinal field curve shows significant evidence for a complex magnetic topology with a strong quadrupolar component. This is supported by the magnetic map for HD~54879 that we present -- the first magnetic map of this star -- which suggests a surface-averaged magnetic field strength of 2954~G. 
The dipole component has a polar strength of 3939~G and is nearly aligned with the rotation axis ($\beta = 168$\textdegree). There is a relatively large quadrupolar component to the magnetic field, with 74\% of the magnetic energy in the dipole, and 23\% in quadrupolar components.  
HD 54879 is indeed one of the most strongly magnetic O-type stars identified to date.
  
During ZDI modelling we found that that the total Gaussian line broadening is 7.2 \kms, including turbulent and thermal broadening. 
This is unusually low for an O-type star \citep[e.g.,][]{Simon2017}. \citet{Sundqvist2013} investigated turbulent broadening in magnetic O stars, finding only NGC 1624-2 with a lower turbulence than we find for HD 54879, while other stars had turbulent broadening of $\gtrsim 20$ \kms. This study relied on known rotation periods from magnetic O stars to remove the ambiguity between rotational and turbulent broadening. They proposed that the very low turbulent velocity in NGC 1624-2 was a result of a strong magnetic field suppressing turbulence.  Additionally, they provide a method for estimating the depth to which a magnetic field may stabilize the atmosphere against turbulence, by looking for the depth where the magnetic and gas pressures are equal.  That depth is parameterized by the local temperature \citep[$T_0$,][Eq.~3]{Sundqvist2013}.  Calculating $T_0$ using $T_{\rm eff}$ and $\log g$ from \citet{Shenar2017} and our mean $|B|$ (2954 G) yields $T_0 = 54 \pm 7$ kK (with the uncertainty dominated by $\log g$).  In the strongest magnetic regions this may rise to $\sim$82 kK.   \citet{Sundqvist2013} propose that the turbulence may originate in the atmosphere between 100 and 200 kK, possibly from an iron opacity bump at $\sim$160 kK.  
\citet{MacDonald2019} investigated this question in more detail, and provided estimates of the magnetic field strength necessary to inhibit convection from this iron opacity bump.  For the parameters of HD 54879, a vertical field of $\sim$5 kG is needed, which is below our mean field but close to the maximum field value.
It appears that the magnetic field of HD 54879 is not sufficient to stabilize the atmosphere down to the iron opacity bump.  
However, comparing with a TLUSTY model atmosphere for $T_{\rm eff} = 30$ kK and $\log g = 4.5$ \citep{Lanz2003}, the magnetically stabilized region of the atmosphere ($T < T_0$) reaches down to an optical depth of $\sim$15.    
Alternatively, HD 54879 is a very cool and relatively high $\log g$ O-type star, so if the generation of turbulence is particularly sensitive to $T_{\rm eff}$ and $\log g$ (or mass and luminosity) in this parameter range, that may also explain the low turbulent velocity.

In parallel, we also investigated how H$\alpha$ emission varies with rotation, finding that the equivalent width of the line varies fairly coherently with $\bz$. The correlation between the magnetic results and the equivalent widths overall supports the interpretation that the variability in \ha is largely driven by a magnetically confined wind. 
We present a fit to the \ha equivalent widths using the rotation period derived from $\bz$.
However, assuming that these variations are rotationally modulated, an independent fit using a 2-component harmonic function yields a different (longer) period, albeit within the same order of magnitude. 
While additional stellar monitoring will be needed to resolve this discrepancy, we conclude that the H$\alpha$ equivalent width variation supports the notion that HD~54879 is an extreme slow rotator: the magnetic O-type star with the second longest rotational period known to this day. 

Some properties of the magnetosphere can be estimated without detailed modeling, although the results are rather approximate for a complex magnetic field and dynamical magnetosphere. 
We calculated the Alfv\'en radius ($R_A$) and wind confinement parameter ($\eta_*$) from \citet{udDoula2008}.
This requires the  mass-loss rate in the absence of a magnetic field ($\dot{M}_{B=0}$), which we estimated from \citet{Vink2001} following their recommendation for the terminal wind speed ($v_\infty$) as 2.6 of the escape velocity \citep{Lamers1995}.
Using $T_{\rm eff} = 30500$~K, $\log L/L_\odot = 4.45$, $R_* = 6.1~R_\odot$, and $M_* = 14~M_\odot$ from \citet{Shenar2017}, yields $\log \dot{M}_{B=0} = -7.9$~$[M_\odot~\textrm{yr}^{-1}]$ and $v_\infty = 2400$~\kms.  
We used the dipole approximation of \citet{udDoula2008} with our dipolar component field strength (3939~G at the pole, 1970~G at the equator), although this neglects some magnetic field strength and complexity, which leads to $\eta_* = 3500$ and $R_A / R_* = 8.0$.  
The Kepler radius \citep{udDoula2008}, calculated with our period, is $R_K/R_* = 310$.
Thus, the star's $R_A$ is large (larger than all but one other O star from \citealt{Petit2013}), but still much smaller than $R_K$, implying that HD~54879 has one of the largest dynamical magnetospheres among O-type stars, and within that rotation is nearly negligible. 

The large Alfv\'en radius will substantially reduce the real mass-loss rate. Realistic calculations accounting for the 3D nature of the magnetic field are non-trivial, and here we limit ourselves to the most commonly used rough approximations\footnote{
An alternative approximate mass-loss rate estimate is provided by \citet{Shenar2017}, who found found $\log \dot{M}_{B=0} \approx 9.0~[M_\odot \textrm{yr}^{-1}]$ and $\log \dot{M} \lesssim 10.2~[M_\odot \textrm{yr}^{-1}]$, illustrating some of the uncertainties involved. }.  
The dipole field approximation from \citet{udDoula2008} for the reduction in mass-loss rate yields $\dot{M}/\dot{M}_{B=0} = 0.09$ ($\log \dot{M} = -8.9 ~[M_\odot \textrm{yr}^{-1}]$).  
The large $R_A$ implies significant angular momentum loss and a relatively short spindown timescale of $\tau_J = \sim 2.6$ Myr \citep{udDoula2009, Petit2013}.  
Combining this with the current very slow rotation of the star implies a spindown age of  $t_J \sim 22$ Myr \citep{Petit2013}.  This is discrepant with the age of 4 or 5 Myr \citep{Castro2015, Shenar2017} from non-magnetic evolutionary tracks.  While the star may have simply reached the main sequence as a very slow rotator, this suggests more sophisticates calculations of the mass-loss and angular momentum loss rates may be needed, accounting for the 3D nature of the magnetic field and wind \citep[e.g.,][]{udDoula2013, Driessen2019b-3D-ADM, Munoz2020, udDoula2023}.

The evolutionary history of HD~54879 has yet to be fully explored.
Previous work by \citet{Castro2015} and \citet{Shenar2017} used the Bayesian statistical tool and method BONNSAI  \citep{2014A&A...570A..66S} to infer the age of HD~54879. This method infers the probability distribution of stellar parameters such as age  via the comparison of observables to any set of stellar evolution models. Both studies used a set of main-sequence, single, non-magnetic massive star evolution models with Milky Way compositions that were produced by \citet{2011A&A...530A.115B}. There was statistical agreement between the inferred ages of $5 \pm 1$ Myr \citep{Shenar2017} and $4.0^{+0.8}_{-1.2}$ \citep{Castro2015}. Additionally, \citet{Shenar2017} inferred an initial mass of $16 \pm 1 M_{\odot}$. However, these models do not account for the effects of magnetic fields such as the reduction of mass loss with respect to evolutionary time scales, so there are added uncertainties that have not been evaluated.
Acknowledging the results of this work, which reveal the complex magnetic topology of the star, suggests that these previous estimates will need to be revised.

\section*{Acknowledgements}

This work is based on observations obtained at the Canada-France-Hawaii Telescope (CFHT) which is operated by the National Research Council (NRC) of Canada, the Institut National des Sciences de l'Univers of the Centre National de la Recherche Scientifique (CNRS) of France, and the University of Hawaii. The observations at the CFHT were performed with care and respect from the summit of Maunakea which is a significant cultural and historic site.

We acknowledge the reverence and importance that Maunakea holds within the Hawaiian community.
Hundreds of historic sites, archaeological remains, shrines, and burials are on its slopes and summit.
CFHT operates on the land of the Native Hawaiian people. We stand in solidarity with Native Hawaiians in their demands for shared governance of Maunakea and to preserve this sacred space for native Hawaiians

The authors also gratefully acknowledge support for this work through the Munich Institute for Astro-, Particle and BioPhysics (MIAPbP) which is funded by the Deutsche Forschungsgemeinschaft (DFG, German Research Foundation) under Germany´s Excellence Strategy – EXC-2094 – 390783311.

CPF acknowledges funding from the European Union's Horizon Europe research and innovation programme under grant agreement No. 101079231 (EXOHOST), and from the United Kingdom Research and Innovation Horizon Europe Guarantee Scheme (grant number 10051045).

ADU acknowledges support from NASA under award number 80GSFC21M0002. SSS acknowledges support from Delaware Space Grant College and Fellowship Program (NASA Grant 80NSSC20M0045).

The authors also thank Dr. Zsolt Keszthelyi for the insights he offered during discussions of this project in the early stages of this manuscript.

The authors also extend their thanks to the anonymous referee, who provided several thoughtful comments that led to the improvement of this work.

\section*{Data Availability Statement}

The data used in this work can be accessed via the public archives maintained by the Canadian Astronomy Data Centre (\esp; \url{https://www.cadc-ccda.hia-iha.nrc-cnrc.gc.ca/en/cfht/}; follow the link and then click ``\esp'' under ``Data'' on the right-hand side of the web page) and by Polarbase (\esp~and Narval; \url{http://polarbase.irap.omp.eu/}).

The authors can also be contacted with any questions about the data included or analyses completed in this work.

\section*{Software Availability Statement}

The spectral normalization performed for this work was computed using the \texttt{normPlot} software package. Least-Squares Deconvolution profiles were calculated using the \texttt{LSDpy} software package. Both \texttt{normPlot} and \texttt{LSDpy} are contained within the \texttt{SpecpolFlow} software package, which is available for installation from \url{https://pypi.org/project/specpolFlow/}. \texttt{SpecpolFlow} is also available via GitHub at \url{https://github.com/folsomcp/specpolFlow}. 
Detailed documentation and tutorials for \texttt{SpecpolFlow}, together with installation instructions, are available on the project's website at \url{https://folsomcp.github.io/specpolFlow/}. The packages are free and open-access, and are licensed under the MIT (\texttt{normPlot}, \texttt{LSDpy}) or GPL-2.0 (\texttt{SpecpolFlow}) open-source software licenses.

\texttt{normPlot} is available for individual installation via the Python Package Index (PyPI) at \url{https://pypi.org/project/normPlot/}. More information on \texttt{normPlot} can be found at \url{https://github.com/folsomcp/normPlot} or on the \texttt{SpecpolFlow} website.

\texttt{LSDpy} is available for individual installation via the Python Package Index (PyPI) at \url{https://pypi.org/project/LSDpy/}. More information on \texttt{LSDpy} can be found at \url{https://github.com/folsomcp/LSDpy} or on the \texttt{SpecpolFlow} website.


\bibliography{ref}{}
\bibliographystyle{aasjournal}

\end{document}